      [1999/12/01 v1.4c Il Nuovo Cimento]
\documentclass{cimento}

\usepackage{graphicx} 
\title{Spectro-photometric study of the GRB 030329 host galaxy\thanks{Based on  
observations carried out with the Nordic Optical Telescope, operated on the
island  of La  Palma  jointly  by Denmark,  Finland,  Iceland, Norway,  and
Sweden,  in the  Spanish Observatorio  del Roque  de los  Muchachos  of the
Instituto de Astrof\'{\i}sica de Canarias. Based on data taken at the 2.2-m
and 3.5-m telescopes of the  Centro Astron\'omico Hispano Alem\'an de Calar
Alto,  operated  by the  Max  Planck  institute  of Heidelberg  and  Centro
Superior  de Investigaciones  Cient\'{\i}ficas.  The  spectral observations
were obtained at the  European Southern Observatory, Cerro Paranal (Chile),
under the Director's Discretionary Time programme 271.D-5006(A).}}
\author{J.~Gorosabel\from{1},
D.~P\'erez-Ram\'\i rez\from{2},
J.~Sollerman\from{3},
\ETC
~A.~de Ugarte Postigo\from{1},
J.P.U.~Fynbo\from{4},
A.J.~Castro-Tirado\from{1},
P.~Jakobsson\from{4},
L.~Christensen\from{6},
J.~Hjorth\from{4},
G.~J\'ohannesson\from{5},
S.~Guziy\from{1},
J.M.~Castro Cer\'on\from{4},
G.~Bj\"ornsson\from{5},      
V.V.~Sokolov\from{7},        
T.A.~Fatkhullin\from{7},     
\atque
K.~Nilsson\from{4}}
\instlist{\inst{1} Instituto de Astrof\'{\i}sica de Andaluc\'{\i}a (IAA-CSIC),
           P.O. Box 03004, E-18080 Granada, Spain
  \inst{2} Universidad de Ja\'en, Departamento de F\'\i sica (EPS),
           Virgen de la Cabeza, 2, Ja\'en, E-23071, Spain
  \inst{3} Stockholm Observatory, Department of Astronomy, AlbaNova, S-106
           91 Stockholm, Sweden
  \inst{4} Niels Bohr  Institute, University of  Copenhagen, Juliane Maries
           Vej  30, 2100  Copenhagen \O,  Denmark
  \inst{5} Science Institute, University of Iceland, Dunhaga 3, 107,
           Reykjav\'{\i}k, Iceland
  \inst{6} Astrophysikalisches Institut, 14482 Potsdam, Germany
  \inst{7} Special Astrophysical Observatory of the Russian Academy of
           Sciences, Nizhnij Arkhyz, 357 147 Karachai-Cherkessia, Rusia
}

\PACSes{\PACSit{95.75.De}{Photography and photometry}
        \PACSit{95.75.Fg}{Spectroscopy and spectrophotometry}
        \PACSit{98.70.Rz}{gamma-ray sources; gamma-ray bursts}}

\begin{document}

\maketitle

\begin{abstract}
  In  this  study  optical/near-infrared(NIR)  broad  band  photometry  and
  optical  spectroscopic observations  of the  GRB 030329  host  galaxy are
  presented.   The  Spectral  Energy  Distribution  (SED) of  the  host  is
  consistent  with a  starburst  galaxy template  with  a dominant  stellar
  population age of  $\sim150$ Myr and an extinction  $A_{\rm v} \sim 0.6$.
  Analysis of the  spectral emission lines shows that the  host is likely a
  low  metallicity  galaxy.  Two  independent  diagnostics,  based  on  the
  restframe UV continuum and the  [\rm OII] line flux, provide a consistent
  unextincted star  formation rate of  SFR$\sim 0.6 M_{\odot}  $ yr$^{-1}$.
  The low absolute magnitude of the  host ($M_B \sim -16.5$) implies a high
  specific   star  formation   rate  value,   SSFR$=  \sim   34  M_{\odot}$
  yr$^{-1}(L/L^{\star})^{-1}$.
\end{abstract}

\section{Introduction}

GRB 030329  was detected on 2003 March  29 at 11:37:14.67 UT  by the HETE-2
spacecraft \cite{Vand03}.   The GRB showed  a duration of  approximately 30
seconds in the  30--400 keV energy range.  Thus,  GRB~030329 falls into the
``long-duration''  category  of   GRBs  \cite{Kouv93}.   The  spectroscopic
observations  performed  for GRB~030329  \cite{Hjor03,Stan03,Soko03,Kawa03}
strongly confirmed previous evidences
\cite{Bloo99,Cast99}  that  long GRBs  are  related  to  Ic supernova  (SN)
explosions.  The redshift of GRB~030329 was determined to be $z=0.168$ from
early  spectroscopy with  the  Very Large  Telescope (VLT,  \cite{Grei03}).
This makes  GRB~030329 the third  nearest burst overall (GRB~980425  is the
nearest  at   $z=0.0085$  \cite{Gala98},  and   GRB~031203  had  $z=0.1055$
\cite{Proc03}). HST  observations showed  that the host  is a  dwarf compact
(full  width half  maximum, FWHM$\sim  0.5^{\prime\prime}$; \cite{Fruc03}).
However, to date  there are no studies on the host  galaxy SED, which gives
relevant information  on the  extinction, dominant stellar  population age,
star formation rate (SFR) and galaxy type.

\section{Observations}

In the present paper we report imaging and spectroscopic observations
of the GRB 030329 host galaxy.

The  photometric  observations  are   based  on  a  number  of  optical/NIR
facilities  in  order to  compile  a  well  sampled SED.   The  $UBVR$-band
observations where performed with the 2.56-m Nordic Optical Telescope (NOT)
equipped with  MOSCA.  Additional  optical observations were  obtained with
the BUSCA camera \cite{Reif00} at the 2.2-m telescope of Calar Alto (CAHA).
BUSCA allows  simultaneous imaging in  four broad optical bands.   The four
channels  (hereafter named $C1$,  $C2$, $C3$  and $C4$)  resemble Johnson's
$UBRI$ bands,  however they do not  correspond to standard  filters so they
need  to be  calibrated  by observing  spectro-photometric standard  stars
\cite{Bohl95}. Additionally  $JHK^{\prime}$-band data  were acquired  with 
the 3.5-m CAHA telescope  equipped with Omega-Prime.  All these optical/NIR
imaging  data were  collected $\sim$1  year after  the GRB  event  when the
optical afterglow (OA) contribution to the total flux was negligible.

The optical spectroscopic data were acquired using VLT equipped with FORS2.
The observations  were conducted with the  300V grism and  an order sorting
filter GG375,  covering the  $\sim 3800 -  8800$~\AA~ wavelength  range.  A
$1.3^{\prime  \prime}$ wide  slit yielded  a spectral  resolution  of $\sim
10.5$~\AA. The VLT observations were performed $\sim$82 days after the GRB,
when  the OA  was still  contributing  substantially to  the measured  flux
($\sim50\%$ according to  \cite{Guzi05}).  Hence, the spectral measurements
overestimated  the host  galaxy continuum  emission, so  the uncontaminated
photometric points had  to be used to estimate the  continuum level. A more
extended description can be found in \cite{Goro05}

\section{Results}

Based  on the  VLT spectroscopic  data  and using  the R$_{23}$  diagnostic
method \cite{Kewl02},  we show  that the GRB\,030329  host is likely  a low
metallicity  galaxy  ($Z=0.004$,  see  also \cite{Soll05}).   From  fitting
synthetic and  empirical SED  templates, we infer  that the host  galaxy of
GRB\,030329  is  most probably  a  starburst  galaxy.   This result  is  in
agreement  with  the conclusions  reported  by  \cite{Chri04}  based on  an
independent host sample,  who found that GRB hosts  correspond to starburst
galaxies in $\sim90\%$ of the cases.

Based  on the  HyperZ public code  \cite{Bolz00},  we derived  a dominant  stellar
population  age   of  $150\pm80$   years  for  a   instantaneous  starburst
episode. However,  an ideal starburst  where all the star  formation occurs
simultaneously  is not  realistic.  Thus,  in  a real  starburst, like  the
GRB~030329  host, although  the  majority of  the  star formation  occurred
$\sim150$  Myr ago, the  star formation  can not  stop sharply.   Thus, the
inferred age of $\sim150$ Myr should be considered as an upper limit of the
actual GRB progenitor age.

\begin{figure}
\begin{center}
\resizebox{8.5cm}{!}{\includegraphics[bb= 172 141 590 665]{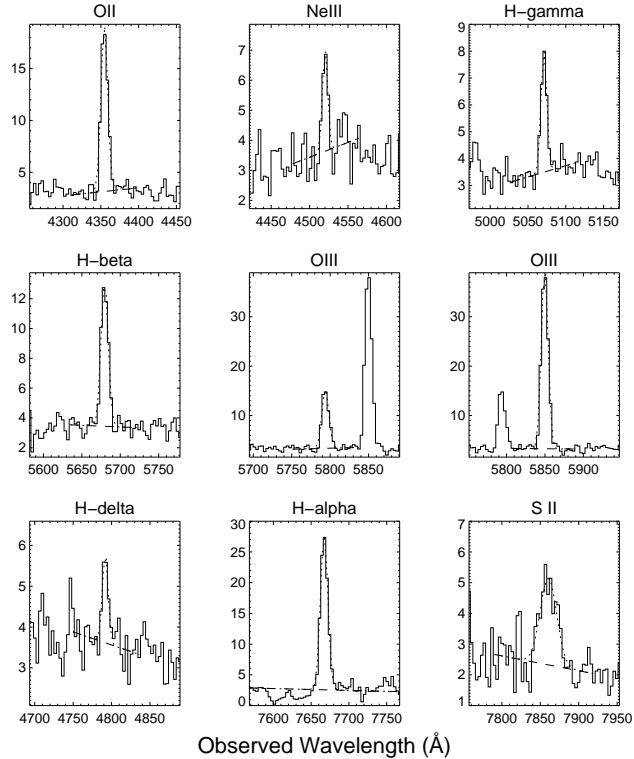}}
\caption{The figure shows the nine lines detected in our VLT spectrum. The
vertical  axes represent  the flux  densities  in units  of $10^{-18}$  erg
s$^{-1}$ cm$^{-2}$  ${\rm \AA}^{-1}$.  The values of  the lines' integrated
fluxes can be seen in \cite{Goro05}.}
\end{center}
\end{figure}

The SED  fitting (using  either synthetic or  empirical templates)  and the
spectroscopic data yield a consistent extinction value of $A_{\rm v} = 0.6$.
Two  independent diagnostic  techniques, namely  the UV  continuum  and the
[OII] emission  line flux,  provide consistently a  SFR value of  $\sim 0.6
M_{\odot}$ yr$^{-1}$ once it is corrected for the host galaxy reddening.

The host restframe $B$-band absolute magnitude is $M_B \sim -16.5$ ($L \sim
0.016 L^{\star}$, assuming $M^{\star}_B=-21$, \cite{Sche76}), confirming its
subluminous nature  \cite{Fruc03}. The GRB\,030329  host SFR is  the lowest
among  the sample  of \cite{Chri04}.   However, the  associated unextincted
specific  SFR  is  the  highest   (SSFR  =  $\sim  34  M_{\odot}$  yr$^{-1}
(L/L^{\star})^{-1}$).   All   the  findings  present  in   this  paper  are
consistent with the host galaxy being an active star forming galaxy.

A GRB\,030329 field galaxy accidentally placed on the FORS2 slit, showed a
redshift very similar to the host galaxy ($z=0.1710\pm0.0003$), which may
be indicative of a possible galaxy clustering. Multi-object spectroscopy of
the GRB\,030329 field might clarify the potential existence of a galaxy
association around $z\sim$0.17

\acknowledgments
This research is partially supported by the Spanish Ministry of Science and
Education   through  programmes   ESP2002-04124-C03-01   and  AYA2004-01515
(including  FEDER funds).  The  observations presented  in this  paper were
partially obtained  under the ESO Programme  271.D-5006(A). JG acknowledges
the support  of a Ram\'on y  Cajal Fellowship from the  Spanish Ministry of
Education  and Science.   The research  of DPR  has been  supported  by the
Education  Council  of Junta  de  Andaluc\'{\i}a,  Spain.   PJ, GJ  and  GB
gratefully  acknowledge support  from a  special grant  from  the Icelandic
Research Council.  VVS and TAF were supported by the Russian Foundation for
Basic Research, grant No 01-02-171061.

\end{document}